\documentstyle[12pt]{article}
\begin{document}

\overfullrule 0 mm \language 0

\vskip 0.3 cm

\centerline { \bf{ DELAYED EQUATION FOR CHARGED}}
\centerline {\bf { RIGID NONRELATIVISTIC BALL }}

  \vskip 1.0 cm

\centerline {\bf{ Alexander A.  Vlasov}}
\vskip 0.3 cm
\centerline {{  High Energy and Quantum Theory}}
\centerline {{Department of Physics}}
\centerline {{ Moscow State University}}
\centerline {{  Moscow, 119899}}
\centerline {{ Russia}}
\vskip 0.3cm
{\it  
Simple expression for self-force acting on radiating rigid charged 
ball is derived (Sommerfeld ball). It is shown that appropriate delayed 
equation of motion has solutions in general differ from that for 
Sommerfeld sphere - there are no "radiationless" solutions, but there are 
oscillating without damping solutions though self-force has nonzero value.}

03.50.De
\vskip 0.3 cm
Long time ago to calculate back-reaction of electromagnetic field on moving 
with acceleration a charged classical (not quantum) particle of finite 
size Sommerfeld [1] considered two models  - uniformly charged  rigid 
sphere and uniformly charged rigid ball and found out explicit expressions 
for electromagnetic self-forces.
  
These expressions in nonrelativistic limit are much simpler then the original 
ones. Thus for uniformly charged rigid sphere (with radius $a$, total 
charge $Q$ and mass $m$ ), moving along trajectory $\vec R = \vec R(t)$, the 
equation of motion with self-force, which include the back-reaction of 
electromagnetic radiation, is the known delayed equation (see, for ex., 
[2-5]) 
 $$m\dot{\vec v}(t) =\vec F_{self} + \vec F_{ext},$$ $$\vec 
F_{self}= \eta \left[ \vec v(t-2a/c) - \vec v(t) \right]\eqno(1)$$ here 
$\vec v = d\vec R /dt$, $\eta\equiv{ Q^2 \over 3\ c a^2 }$, and $\vec 
 F_{ext}$ - some external force.  

This equation 
is often considered in literature (see, for ex., [2-5, 6, 7]) and has the 
following significant features:

1.) there are no "runaway" solutions - all solutions of (1) for zero 
external force exponentially tend to zero for $t \to \infty$ (are damped);

2.) there are "radiationless" solutions - solutions with zero 
value of the self-force, though the body moves with acceleration. In 
particular, for harmonic oscillations appropriate radiationless 
frequencies have discrete values.

As for charged rigid ball with radius $a$ and total charge $Q$, till now 
there were not known the compact forms of nonrelativistic self-force and 
the following expansion in powers of $a/cT$ was used:
 $$\vec 
 F_{self}(t) =-{24 Q^2\over a c^2 } \cdot \sum_{n=o}^{\infty} {(-2a)^n \over 
c^{n} n!(n+2)(n+3)(n+5)}\cdot {d^{n+1} \over d t^{n+1}}\vec 
v(t) \eqno(1)$$ 

One can derive this result as from the strict Sommerfeld result so from 
the well-known general Jackson form for a self-force in nonrelativistic, 
linear in velocity and its derivatives approximation for spherically 
symmetrical body [8]:  $$\vec 
 F_{self}(t) =-{2 \over 3\ c^2 } \cdot \sum_{n=o}^{\infty} {(-1)^n \over 
c^{n} n!}\cdot \gamma_{n-1}\cdot{d^{n+1} \over d t^{n+1}}\vec v(t) \eqno 
(2)$$ 
here $\gamma_{n-1}$ - are the  form-factors, do not depending from the 
time in approximation under consideration:  $$\gamma_{n-1}= \int d\vec r 
  d\vec r' \   \rho\ \rho'\ |\vec r - \vec r'|^{n-1}\eqno(3)$$

The density of the charge $\rho$ for rigid uniformly charged ball is 
$$\rho(\vec r)=\rho_0 \cdot \theta(a-r),\ \ \ 
\rho_0\equiv {Q\over 4\pi a^3/3} \eqno(4)$$ 
With this density the form-factors 
$\gamma_{n-1}$ in (2) are calculated as  
$$\gamma_{n-1}=\int d\vec r d\vec r' \ \rho\ \rho'\  |\vec r - \vec 
r'|^{n-1}={36 Q^2 \over a}\cdot {(2a)^n \over (n+2)(n+3)(n+5)}\eqno(5)$$ 
Using (5) in (2), we get the result (1).

For our goal - to construct compact form of self-force in 
case under consideration - it will be more convenient to 
use another representation of Jackson result (2-3).

It is not too complicate to notice that the expansion (2-3) can be 
rewritten in the form of Taylor expansion of delayed acceleration $\vec a 
= \dot {\vec v}$  of a body:  $$  \vec {F}_{self}(t) = -{2 \over 3  c^2 } 
\int \int d\vec r d\vec r'{\rho \rho' \over |\vec r - \vec r'|} 
\sum_{n=o}^{\infty} {1 \over n!} \left({-|\vec r - \vec r'| \over c}\cdot 
{d \over dt}\right)^n  \vec a (t) =$$ $$=- {2 \over 3  c^2 } \int \int 
d\vec r d\vec r'{\rho \rho' \over |\vec r - \vec r'|} \vec a (t -{ |\vec r 
- \vec r'| \over c})\eqno(5)$$ After substitution of $\rho$ in form of (4) 
 into (5) and integration, the expression (5) reduces to $$ \vec 
{F}_{self}(t)= -k \bigg[{2 \over 3} ({a \over c})^3 \vec {R}(t) - ({a 
\over c})^2 \left(\vec {R} ^{[-1]} (t) +\vec {R} ^{[-1]} (t-2{a \over c}) 
\right)-$$ $$- 2{a \over c} \vec {R} ^{[-2]} (t-2{a \over c})  + \vec {R} 
^{[-3]} (t) -\vec {R} ^{[-3]} (t-2{a \over c}) \bigg]  \eqno(6) $$ here 
 $$\vec {R} ^{[-n]} (t)\equiv \int\limits^{t} dt_1 ...\int 
\limits^{t_{n-1}}dt_n \vec {R} (t_n) $$ and $$k \equiv {3Q^2c^3 \over  
a^6} $$
The equation of motion of Sommerfeld ball then reads:
$$\ddot {\vec R}(t) = \vec {F}_{self}(t) /m + \vec {F}_{ext}(t)/m \eqno(7)$$

To compare features of equation (7) with that for Sommerfeld sphere we 
consider the case of zero external force and search for solutions of (7) 
in the form of harmonic oscillations with frequency $w$, in general 
being complex number:
$$ \vec R(t)^{[-3]} = \vec A \exp{(iwt)} \eqno (8)$$
Consequently eq.(7) reduces to the transcendental equation which for new 
dimensionless variable $W \equiv iw\delta\ \ \ (\delta\equiv 2a/c)$ is
$${1\over k^{*} } W^5 =-{1 \over 12} W^3 +{1 \over 4} W^2 -1 
+\bigg(1+{1\over 2} W\bigg)^2 \exp{(-W)} \eqno(9)$$
here $k^{*} \equiv {3 Q^2 c^3 \over m a^6}\cdot {(2a)^5 \over c^5} ={96 
Q^2\over mc^2 a}$ - is dimensionless parameter.

R.H.S. of (9) is proportional to the value of the self-force (6), 
calculated for harmonic oscillations (8).

 Consider the following cases.

1.

Let $Im (W) =0$, i.e. the frequency $w$ is purely imaginary: $w=iz$.

Then the function in R.H.S. of eq. (9) has zero value for $W=0$ and for 
$W>0, \ \ (z<0)$ its derivative with respect to $W$ has negative sign:
$${d\ R.H.S \over dW} =W/2\bigg(1-W/2- (1+W/2)\exp{(-W)}\bigg) <0$$
Thus this function is negative for $W>0$. Consequently the equation (9), 
where L.H.S. is positive for $W>0$, has no solutions for $W>0,\ \ (z<0)$. 
Thus there are no solutions for $R(t)$ of the form $$ R \sim 
\exp{(-z t)},\ \ z<0$$
 - i.e there are no exponentially  increasing with time $t$ solutions.
 
So for Sommerfeld ball as for Sommerfeld sphere there are no exponentially 
"running away solutions". 

2.

Let $Re (W) =0,\ \ \ W\equiv iZ$, i.e. the frequency $w$ is purely real 
number.

Then eq. (9) can be put in the form
$$iZ^5/k^{*}-iZ^3 /12+Z^2/4+1 =\exp{(-iZ)}(1+iZ/2)^2 \eqno (10)$$
Multiplication of (10) on its complex conjugate leads to the relation
$$(Z^5/k^{*} -Z^3/12)^2=0 \eqno (11)$$
It has the solution
$$Z=\pm \sqrt{k^{*} /12} \eqno(12)$$
Turning back to the eq.(10) with (12), we come to the result
$$Z^2/4+1 = \exp{(-iZ)}(1+iZ/2)^2\eqno(13)$$
Reducing the R.H.S. of (13) to the form
$$(Z^2/4+1)\exp{(-iZ+i 2 \Phi)},\ \ \ tg (\Phi) =Z/2$$
we get from (13) 
$$Z = 2\Phi,\ \ or \ \ tg(Z/2) =Z/2 \eqno(14)$$
So relations (14) and (12) taking together yield the solution of our 
problem:

if $Z=Z_{n},\ \ n=1,2,3,...$ - the discrete solutions of the 
transcendental eq. (14), then parameter $k^{*}$ must satisfy the condition
$$ {k^{*}\over 12} \equiv {8 Q^2  \over mc^2 a} = (Z_{n})^2 $$

Thus for these frequencies (and values of parameter $k^{*}$ ) 
there are harmonic oscillating solutions of  ball equation (7), free of 
radiation damping. 

This class of solutions is absent for Sommerfeld sphere.

On the other hand, taking
$$Z^5/k^{*} \to 0 \eqno(15)$$
one can try to find weather there are the radiationless solutions ( 
for which $\vec F _{self} =0$  or  $\bigg(R.H.S.\  of\  (9)\bigg)_{W=iZ} 
=0 $) for the ball equation (7).  Equation  (11) due to (15) 
gives the immediate answer - there are no such solutions.

Thus one can see that Sommerfeld ball  and Sommerfeld sphere models
provide us with  different behavior of moving charged body. For the 
Sommerfeld ball there are no "radiationless" solutions but  there are 
solutions with not damped oscillations though the value of the self-force 
differs from zero.

\eject

\centerline {\bf{REFERENCES}}

  \begin{enumerate}
\item A.Sommerfeld, Gottingen Nachrichten, 29 (1904), 363 (1904), 201 
  (1905).
\item T.Erber, Fortschr.Phys., 9, 342 (1961).

\item P.Pearle, in {\it Electromagnetism}, ed. D.Tepliz, Plenum, NY,
1982, p.211.

\item S.Parrott, {\it Relativistic Electrodynamics and Differential
Geometry },  Springer-Verlag, NY, 1987.

\item F.Rohrlich, Am.J.Physics, 65(11), 1051 (1997), Phys.Rev., D60, 084017 
(1999) 

\item Al. A. Vlasov, Vestnik Mosk. Univer., Fizika, N 5, 17 (1998), N 6, 
15 (2001)
 \item Alexander. A.Vlasov, in "Photon: old problems in light of 
new ideas", p.  126, ed. Valeri V. Dvoeglazov, Nova Sci. Publ., NY, 2000.  
 E-print Archive: physics/9911059, 9912051, 
0004026, 0103065, 0110034.  
\item J. D. Jackson, {\it Classical Electrodynamics}, 3rd ed., Wiley, NY, 
1999.

   \end{enumerate}

\end{document}